\documentclass[prc,preprint]{revtex4-1}

\usepackage{amsmath}
\usepackage{amsfonts}
\usepackage{graphicx}
\usepackage{multirow}

\begin{document}
\title{Neutrino oscillations: ILL experiment revisited}
\author{B. K. Cogswell}
\affiliation{School of Physics and Astronomy, University of Manchester, Manchester M139PL, United Kingdom}
\author{D. J. Ernst}
\affiliation{Department of Physics and Astronomy, Vanderbilt University, Nashville, Tennessee 37235}
\author{J. T. Gaglione}
\affiliation{Math and Natural Science Division, Nashville State Community College, Nashville, Tennessee 37209}
\author{K. T.  L. Ufheil}
\affiliation{Department of Physics and Astronomy, Vanderbilt University, Nashville, Tennessee 37235}
\author{J. M.  Malave}
\affiliation{Department of Physics and Astronomy, Baldwin Wallace University, Berea, Ohio 44017}

\date{\today}
\begin{abstract}
The ILL experiment, one of the ``reactor anomaly'' experiments, is re-examined. ILL's baseline of 8.78 m  is the shortest of the reactor anomaly short baseline experiments, and it is the experiment that finds the largest fraction of the electron antineutrinos disappearing -- about 20\%.  Previous analyses, if they do not ignore the ILL experiment,  use functional forms for chisquare which are either totally new and unjustified, are the magnitude chisquare (also termed a ``rate analysis''), or utilize a spectral form for chisquare which double counts the systematic error. We do an analysis which utilizes the standard, conventional form for chisquare as well as a derived functional form for a spectral chisquare. We find that when analyzed with a conventional chisquare that includes spectral information or with a spectral chisquare that is independent of the magnitude of the flux, the results are a set of specific values for possible mass-squared differences of the fourth neutrino and where the  minimum chisquare difference values are significantly enhanced over previous analyses of the ILL experiment. For the Huber flux and the conventional chisquare, the two most preferred values are mass-squared differences of 0.90 and 2.36 eV$^2$ preferred at  $\Delta \chi^2_{min}$ values of -12.1 and -13.0 (3.5 and 3.6 $\sigma$), respectively. For the Daya Bay flux and conventional chisquare we find 0.95 and 2.36 eV$^2$ preferred at $\Delta\chi^2_{min}$ of -8.22 and -9.45 (2.9 and 3.1 $\sigma$), respectively. For the spectral chisquare and either flux these values are 0.95 and 2.36 eV$^2$ preferred at $\Delta\chi^2_{min}$ of -10.5 and -11.7 (3.2 and 3.4      $\sigma$), respectively. These are to be compared to -4.4 (2.1 $\sigma$) found in the original reactor anomaly anaysis for all of the experiments excepting the ILL experiment.
\end{abstract}
\pacs{???}

\maketitle
\section{Introduction}
Neutrino oscillation experiments have, over the last decade, moved toward precision measurements \cite{sala,este,capo}. A principle goal of these experiments is to determine the six phenomenological mixing parameters --- three mixing angles, $\theta_{12}$, $\theta_{13}$, and $\theta_{23}$; two mass-squared differences $\Delta m^2_{21}$ and $\Delta m^2_{31}$; and the CP violating phase $\delta$. The mixing angles, $\sin^2\theta_{12}$, $\sin^2\theta_{13}$, and $\sin^2\theta_{23}$ are found \cite{sala} to be $3.20^{+0.20}_{-0.16} \times 10^{-1}$, $2.160^{+0.083}_{-0.069}\times 10^{-2}$ ($2.220^{+0.074}_{-0.076} \times 10^{-2}$), and $5.47^{+0.20}_{-0.30}\times 10^{-1}$ ($5.51^{+0.18}_{-0.30}\times 10^{-1})$ respectively, where the hierarchy is given by normal (inverse).  The mass squared-differences $\Delta^2_{21}$ and $\vert\Delta^2_{31}\vert$ are found to be $7.55\pm 0.03\times 10^{-5}$ and  $2.50\pm0.03\times 10^{-3}$ ($2.42^{+0.03}_{-0.04}\times 10^{-3})$ eV$^2$, respectively. Note that the errors range from just under 2\% to 6\%, defining our new precision era. There is also evidence \cite{sala,este,capo} at the 1 to 2 $\sigma$ level indicating a non-zero value for $\delta$, with its preferred value being near 3$\pi$/2. Only a small indication of which hierarchy is correct is found.

However, there are experiments that are not consistent with the three neutrino analyses. These experiments require a mass-squared difference of order 1 eV$^2$. These experiments are:
\begin{itemize}
\item The LSND \cite{LSND} and MiniBoone \cite{mini} experiments measure $\overline\nu_\mu\rightarrow\overline\nu_e$ and $\nu_\mu\rightarrow\nu_e$ oscillations. The LSND experiment indicates a sterile neutrino that oscillates via a mass eigenstate with a mass-squared difference that is greater than 0.1 eV$^2$.  MiniBoone has a longer baseline and compensating larger energy than LSND. These two experiments have recently been found \cite{MB} to be compatible. 
\item Experiments with radioactive sources at the Gallium solar neutrino facilities, Sage and Gallex \cite{Giu}, see fewer neutrinos than expected. This can be explained by the disappearance of electron neutrinos oscillating via a mass-eigenstate with  mass-squared difference greater  than 1 eV$^2$. This is called the ``gallium anomaly''.
\item A new calculation of the electron antineutrino flux \cite{Mue} yielded a net increase of the predicted rate of antineutrinos emitted by the four dominant decays that drive a reactor. This implied \cite{Ment} for a number of short-baseline reactor experiments from the 1980's and 1990's that the antineutrinos oscillated away via a mass-eigenstate with $\Delta m^2 >$ 1 eV$^2$ . This is called the ``reactor anomaly''.
\item There are five recent reactor antineutrino oscillation experiments: Nucifer \cite{Nuc}, NEOS \cite{NEOS}, Neutrino-4 \cite{N-4}, DANSS \cite{DANS}, and PROSPECT \cite{PROS}. A flux independent analysis of \cite{NEOS} and \cite{DANS} combined with the Gallium \cite{Giu} experiments is presented in Ref.~\cite{Gari}. 
\end{itemize} 

This work focuses on one of the reactor anomaly experiments -- the ILL experiment \cite{kwon,houm}. This experiment is distinctive in several ways. It has an 8.79 m baseline, the shortest baseline of any of the reactor anomaly experiments.  The short baseline gives ILL sensitivity to the largest values for $\Delta m^2$.  The original publication \cite{kwon} of this experiment found the total number of measured antineutrinos to be  4.5 $\pm$ 11.5\% less than predicted. However, the power of the reactor was found \cite{houm} to have been under-measured by 18\%, implying that approximately 20\% of the antineutrinos had disappeared. This is by far the largest fraction of electron antineutrinos  disappearing in any  short-baseline reactor experiment. 

In the Mention analysis \cite{Ment}, the reactor anomaly data indicate that a fourth antineutrino exists at the 2.1 $\sigma$ level, but the ILL experiment is omitted from this analysis. When they combine other data with the reactor anomaly data, they use a spectral chisquare function for the ILL experiment which we argue below is incorrect. In the Kopp-Dentler  analysis \cite{Kopp,Dent,Dent2}, the magnitude chisquare is used for all but the Bugey experiment \cite{bug}. Use of the magnitude chisquare  underestimates the impact of experiments which have spectral information, including ILL. They find that the reactor anomaly experiments indicate the existence of a fourth antineutrino at the 2.7 $\sigma$ level. In the Collin analysis \cite{Coll}, only the Bugey \cite{bug} experiment from the reactor anomaly experiments is included. In the Gariazzo study \cite{Gari2} only the magnitude analysis for the ILL experiment is used. They find a 2.9 $\sigma$  indication of a fourth antineutrino after including results from the NEOS experiment, and the near detector data from the Daya Bay \cite{dbay}, RENO \cite{RENO}, and Double CHOOZ \cite{DCh} experiments were also included. They find that the existence of a fourth antineutrino is indicated at a 3.1 $\sigma$ level when these additional data are included. There is agreement that evidence exists supporting the existence of a fourth neutrino, but a correct analysis of the ILL experiment beyond the use of the magnitude chisquare (a rate analysis) does not yet exist. 

Here, we address two fundamental questions within the context of providing new results for the ILL experiment. The first, in Sections II and III, the importance of the choice of the chisquare function used in the analysis is examined. The second, in Section IV, the dependence of the results on the choice of the flux is presented.  We find that including the spectral information gives results that favor a number of particular values of $\Delta m^2$. In Section V we demonstrate how this comes about. In Section VI, we review our conclusions and comment on possible future work.
\section{CHISQUARE FUNCTIONS}

Given that different authors utilize different functional forms for their chisquare function, we ask the question of how does the choice of the chisquare function impact the physical results implied by an analysis of the experiment? We give the formulae for each of the chisquare functions of interest. We postulate that one is not free to create any function one chooses. It is necessary to extract from the calculated chisquare the answer to various questions that involve probabilities. This usually is done by knowing that the likelihood function that results from the chisquare function is a probability distribution. To be correct, a mathematical proof of how to extract probabilities is required. Here, we maintain this constraint by limiting ourselves to the conventional chisquare and normal (Gaussian) statistics. This is the standard chisquare found in books on probability theory. The extraction of probabilities then follows a prescription which has been rigorously derived, using what is commonly called a ``frequentist" approach or else a ``Bayesian" \cite{Bur,Ber} approach. One can divide this conventional chisquare into two parts. One part we call the spectral chisquare. This chisquare is independent of the magnitude of the antineutrino flux. This second form is the limit in which one simply counts the number of neutrinos without measuring their energy. This is also called a ``rate" calculation. This is the limit of the conventional chisquare when there is only one energy bin. Since both of these chisquares derive from the conventional chisquare, extracting probabilities utilizing either the ``frequentisit" or ``Bayesian" approach is mathematically rigorous. In addition, we examine the results of using the sum of the magnitude and spectral chisquare. The sum of the two parts is not rigorously equivalent to the conventional chisquare. 

We begin with the well-known and mathematically rigorous conventional $\chi^2$ function as given by  
\begin{eqnarray}
\chi^2_{\rm conv}(\sin^2 2\theta ,\,\Delta m^2)&=&\sum_{i=1}^{i_{max}}\,\frac{(N^{\rm exp}_i-N^{\rm th}_i(\{a\},E_i,\sin^2 2\theta, \Delta m^2 ))^2 }{(\sigma_i\, N_i^{\rm exp})^2}\nonumber\\
&&+ \sum_{j=1}^{j_{max}}\,\frac{(a_j-1)^2}{\hat\sigma_j^2}\,,
\label{eq1}
\end{eqnarray}
where $N^{\rm exp}_i$ is the experimentally measured number of neutrinos and $\sigma_i$ is its statistical error given in percent of $N^{\rm exp}_i$, both for bin $i$ centered at energy $E_i$, with $i_{max}$ being the total number of spectral bins. We stress that this functional form of the chisquare follows in a mathematically rigorous way from the property that the data satisfies normal statistics. By dividing the number of neutrinos by the run time, the number of neutrinos can be replaced by the rate of measuring the neutrinos in all formulae. Systematic errors are included through the use of a set of nuisance parameters $\{a\}$  with $j_{max}$ the number of such parameters.  These parameters are varied, subject to the constraint imposed by $\hat\sigma_j$ in the formula.  $N^{\rm th}_i(\{a\},E_i,\sin^2 2\theta, \Delta m^2 )$ is the theoretical model for the number of neutrinos in bin $i$. We use a two neutrino model, with our independent variables taken as $\sin^2 2\theta$ and $\Delta m^2$. The two neutrino approximation results \cite{Coll} from taking $\Delta m^2_{21}=\Delta m^2_{32}=0$ in the full four neutrino mixing matrix. 

The probability that an electron antineutrino leaving the reactor remains an electron antineutrino when it arrives at the detector is given by
\begin{equation} 
{\mathcal P}_{ee}(L,E, \sin^2 2\theta,\Delta m^2)=1-\sin^2 2\theta\,\sin^2(1.267\, \Delta m^2\,L/E)\,\,,
\label{eq2} 
\end{equation}
where $L$ is the distance traveled by the antineutrino in m and $E$ is its energy in MeV. The mass-squared difference parameter is in units of eV$^2$. In order to incorporate the finite energy resolution of the detector the oscillation probability must be convoluted with an energy resolution function, $f(E-E')= N_E\,exp(-(E'-E)^2)/2\sigma_E^2)$, with $N_E$ a normalization factor.  The distance $L$ must be averaged over the distance between points in the core and points in the detector. This can be done with a one dimensional integration by defining a weight function $W_L(L')/L'^2\,dL'$ that extends from the smallest (largest) distance, $L_{min}$ ($L_{max}$) between a point in the core to a point in the detector. We divide $L_{max} - L_{min}$  into bins. The $1/L'^2$ factor accounts for the inverse square drop in the flux with distance. We generate randomly located pairs of points with constant density in the core and in the  detector and calculate the distance between each pair of points, then put a point in the appropriate bin for that distance. The number of points in each bin then gives a weight function, $W_L(L')/L'^2\,dL'$, which we normalize. With this weight function, we  then need only do a one dimensional integral over $L'$ weighted by $W_L(L')/L'^2$.  To include these two effects, we define this averaged $P_{ee}$ by
\begin{eqnarray}
<{\mathcal P}_{ee}&&(L,E, \sin^2 2\theta,\Delta m^2) > \,=\cr
&& N_E\,\int^{ \infty}_{E_{th}}\,dE' \,F(E-E')\,\int_{Lmin}^{Lmax}dL'\,W_L(L')/L'^{2}\,\,{\mathcal P}_{ee}(L',E', \sin^2 2\theta,\Delta m^2)\,\,,
\label{eq3}
\end{eqnarray}
where $E_{th}$ is the antineutrino threshold energy for the inverse beta decay reaction. The theoretically predicted number of neutrinos in bin $i$ is  then 
\begin{equation}
N^{\rm th}_i(a, \sin^2 2\theta, \Delta m^2 ) = a\,N^{\rm no}_i(E_i)\,<{\mathcal P}_{ee}(L, E_i, \sin^2 2\theta, \Delta m^2)>\,,
\label{eq4}
\end{equation}
where  $N_i^{\rm no}$ is the theoretical number of neutrinos that would have been measured in bin $i$ in the absence of oscillations, and $a$ is the one nuisance parameter we employ.

An alternative approach, as used in Ref.~\cite{houm}, arises from separating the $\chi^2$ function into two pieces, a magnitude piece, $\chi^2_{\rm mag}$, and a spectral piece, $\chi^2_{\rm spec}$
\begin{equation}
\chi^2_{m+s} \rightarrow \chi^2_{\rm mag} + \chi^2_{\rm spec}\,.
\label{eq5}
\end{equation}
The magnitude part, $\chi^2_{\rm mag}$, describes experiments where the total number of antineutrinos is detected but their energies are not measured. This chisquare function, $\chi^2_{\rm mag}$, is given by
\begin{equation}
\chi^2_{\rm mag}(\sin^2 2\theta, \Delta m^2)  = \frac{{( N^{\rm exp}_{\rm tot}  - N^{\rm th}_{\rm tot}}(a, sin^2 2\theta, \Delta m^2))^2}{(\sigma_{\rm tot}\,N^{\rm exp}_{\rm tot})^2}+\frac{(a-1)^2}{{\hat\sigma}^2}\,,
\label{eq6}
\end{equation}
where $N^{\rm exp}_{tot}$ is the total experimental number of antineutrinos detected, and $N^{th}_{ tot}$ is the total theoretically predicted number of antineutrinos. This is simply a one energy bin form of Eq.~\ref{eq1}.  

The spectral chisquare function, $\chi^2_{\rm spec}$, is constructed to be a chisquare that is independent of the magnitude of the flux. Physically this means you have no knowledge of the magnitude of the flux. This can be accomplished by letting $\hat\sigma$ go to infinity in Eq.~\ref{eq1} yielding 
\begin{equation}
\chi^2_{spec}(\sin^2 2\theta, \Delta m^2) = \sum_{i=1}^{i_{max}} \frac{(N^{\rm exp}_i - N_i^{th}(a, E_i,  \sin^2 2\theta, \Delta m^2))^2}{(\sigma_i\,N^{exp}_i )^2}\,.
\label{eq7}
\end{equation}
In Ref.~\cite{houm} a different and unique form for the spectral chisquare was proposed. We disregard that definition. Our definition is manifestly and completely independent of the flux. In Mention, Ref.~\cite{Ment}, another chisquare function that is independent of the magnitude of the flux is used, a flux we will call $\Delta\chi^2_{specM}$. There they take the limit  of $\hat\sigma$ going to infinity and then insert the systematic error by adding it in quadrature to the statistical error.  This chisquare, $\chi^2_{specM}$, is defined by Eq.~12 in Ref.~\cite{Ment} as
\begin{equation}
\chi^2_{specM}(\sin^2 2\theta, \Delta m^2) = \sum_{i=1}^{i_{max}} \frac{(N^{\rm exp}_i - N_i^{th}(a, E_i,  \sin^2 2\theta, \Delta m^2))^2}{(\sigma_i\,N^{exp}_i )^2+(\hat\sigma\, N^{exp}_i)^2}\,.
\label{eq8}
\end{equation} 
In taking the limit of $\hat\sigma$ going to infinity, the effect of the systematic error has been included; one could even say ``we have over-included it". Putting the second term in the denominator of Eq.~\ref{eq8} is double counting the systematic error. 
 
We have sufficient data to calculate each of these chisquares. In Table IV of Ref.~\cite{kwon} we are given the energy grid, $E_i$; the experimental rate of particles being detected in units of MeV$^{-1}$h$^{-1}$ in each bin; and its error $\sigma_i$.  We use one nuisance parameter with error $\hat\sigma$ for the error in the magnitude of the flux, number of protons, etc.  We find for the error  a value of 11\% in \cite{kwon},  a value of 8.87\% in \cite{houm} , and a value of 9.5\% in \cite{Ment} . We choose to be conservative and use the largest of these, 11\%.  Also from Ref.~\cite{kwon} we get the dimensions needed to construct $W_L(L')$. The core has a radius 0.2 m and a height of 0.8 m. The detector is 1.2 m tall, 0.8 m wide, and (we estimate) 0.9 m deep, the first two taken from the diagram in Fig.~1, Ref.~\cite{kwon}, while the depth is estimated to be 1.0 m, as it is not provided anywhere. We find that the inclusion of the energy resolution integration is not needed as its impact is less than one percent. On the other hand the spatial integration over the size of the core and the size of the detector is approximately a 25\% correction.
\section{Results - chisquare  dependence}

$\chi^2(\sin^22\theta, \Delta m^2)$ does not return to zero as $\Delta m^2$ tends to infinity. It approaches a $\Delta m^2$ independent valley arising from the limit of $\Delta^2 m$ large,
\begin{equation}
\sin^2\left(\frac{1.267\,\Delta m^2\,L}{E}\right) \rightarrow 1/2\,\,.
\end{equation}
The usual approach to extract probabilities from a chisquare function is to define a likelihood function, ${\mathcal L}(\sin^2 2\theta,\,\Delta m^2) =: \exp(-\chi^2(\sin^22\theta,\,\Delta m^2)/2$, and realize that the likelihood function is proportional to a probability distribution. This cannot be done here since the probability distribution is not integrable. The solution to this situation can be found in Ref.~\cite{bug}. The question one asks must be altered and  the approach is termed  the ``raster" interpretation. In this approach, one asks the question ``for a given $\Delta m^2$, what is the minimum (best fit) value of the chisquare and at what value of $\sin^22\theta$ does it occur?'' We define the answer to this question as $\Delta \chi^2_{min} (\sin^2(2\theta_{min}), \Delta m^2)) = \chi^2(\sin^2(2\theta_{min},)\Delta m^2) - \chi^2(0,0)$, where $\theta_{min}$ is the minimum value for the chosen value of $\Delta m^2$. The no oscillation chisquare in the two neutrino analysis is the value of the chisquare function for three neutrinos. Thus $\Delta \chi^2_{min}$ tells you how much better a fit the inclusion of a fourth neutrino yields. Note that the sign of $\Delta \chi^2_{min}$ is the opposite of the sign often used. We use  this sign as the best fit is then given by the smallest value of $\Delta \chi^2_{min}$. Also note that the chisquare is a one variable, $\sin^2(2\theta)$, chisquare, and hence for a frequentist analysis  the improvement due to the fourth neutrino as measured in number of standard deviations is the square root of $\Delta\chi^2_{min}$.  
\begin{figure}
\includegraphics[width=4.2in]{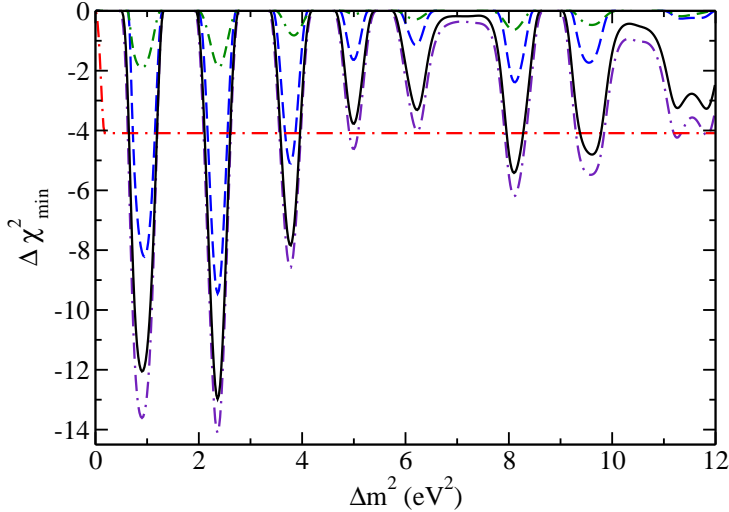}
\caption{$\Delta\chi^2_{min}$ curves versus  $\Delta m^2$ for a variety of choices all using the Huber \protect\cite{Hube} flux. The solid (black) curve is the result for the conventional chisquare, $\chi^2_{conv}$, Eq.~\ref{eq1}; the dot-dash (red) curve is the result for the magnitude or rate chisquared, Eq.~\ref{eq6}; the dash  (blue) curve is the result of  the spectral chisquare, $\chi^2_{spec}$, Eq.~\ref{eq7}; the dot-dash-dash (indigo) curve is the results for the sum of the conventional chisquare and our version of the spectral chisquare, $\chi^2_{m+s}$, Eq.~\ref{eq5}; and the dot-dot-dash (green) curve is the result of the spectral chisquare proposed by Mention, $\chi^2_{specM}$, Eq.~\ref{eq8}. Note that the spectral chisquare, the dash (blue) line, is also the result for the Daya Bay flux.}
\label{fig1}
\end{figure}
\begin{table}
\begin{center}
\begin{tabular}{|c|c|c|c|c|}
\hline
$\Delta \chi^2_{min}$&$~\sin^22\theta_{min}$~~&~~$\Delta m^2_{min}$ (eV$^2$)~~&$~\Delta\chi^2_{min}~$&$~~ \sigma~~$\\
\hline
\multirow{8}{*}{~Conv~}&$0.259$&$0.90$&$-12.1$&3.5\\ \cline{2-5}
                    &$0.267$&$2.36$&$-13.0$&3.6\\ \cline{2-5}
                    &$0.225$&$3.78$&$-7.84$&2.8\\ \cline{2-5}
                    &$0.173$&$5.00$&$-3.77$&1.9\\ \cline{2-5}
                    &$0.187$&$6.23$&$-3.32$&1.8\\ \cline{2-5}
                    &$0.269$&$8.10$&$-5.42$&2.3\\ \cline{2-5}
                    &$0.285$&$9.61$&$-4.81$&2.2\\ \cline{2-5}
                    &$0.303$&$11.3$&$-3.25$&1.8\\ \cline{2-5}
                    &$0.319$&$11.8$&$-3.28$&1.8\\ \hline
\multirow{8}{*}{~Spect~}&$0.233$&$0.95$&$-8.22$&2.9\\ \cline{2-5}
                    &$0.245$&$2.36$&$-9.45$&3.1\\  \cline{2-5}
                   &$0.195$&$3.78$&$-5.10$&2.3\\ \cline{2-5}
                    &$0.127$&$5.00$&$-1.64$&1.3\\ \cline{2-5}
                    &$0.123$&$6.20$&$-1.55$&1.2\\ \cline{2-5}
                    &$0.207$&$8.12$&$-2.39$&1.5\\ \cline{2-5}
                    &$0.199$&$9.54$&$-1.73$&1.3\\ \cline{2-5}
                   &$0.123$&$11.3$&$-0.26$&0.5\\ \cline{2-5} 
                   &$0.115$&$11.7$&$-0.24$&0.5\\ \hline 
\end{tabular}
\caption{ The location of the minima, $\sin^22\theta_{min}$ and $\Delta m^2_{min}$, and the depth of the minima, $\Delta\chi_{min}$ for the conventional chisquare, Eq.~\protect\ref{eq1} and for the spectral chisquare, Eq.~\protect\ref{eq7}. For the conventional chisquare the results are for the Huber flux, while for the spectral chisquare, the results apply to both the Huber flux and the Daya Bay flux.}
\label{tab1}
\end{center}
\end{table}

We first investigate the dependence of results on the choice of the chisquare function in Fig.~\ref{fig1} and in Table~\ref{tab1} we quantify our results by giving the depth for each minima, $\chi^2_{min}$ and its location $\sin^2 2\theta_{min}$ and $\Delta m^2_{min}$.  For all curves in this section we use the Huber \cite{Hube} flux. In Fig,~\ref{fig1} the solid (black) curve depicts  $\Delta \chi^2_{min}$ as a function $\Delta m^2$ for the conventional chisquare, $\chi^2_{conv}$, defined in Eq.~\ref{eq1}.  The first thing we note is that the curve is a set of individual minima. The origin of multiple minima will be investigated in Section V. This phenomenon is new to this work. Each value for the minima is exceptionally deep. The depth of the first two minima are  $\Delta \chi^2_{min}$  = -12.1 and -13.0  (3.4 and 3.6 $\sigma$) and are located at $\Delta m^2 =   $ 0.9 and 2.4 eV$^2$. The result obtained in the Mention work \cite{Ment} is -4.45 (2.1 $\sigma$) for all the reactor anomaly experiments except ILL. The ILL experiment is thus the dominant experiment of the reactor anomaly experiments. This is not surprising since the ILL experiment finds about 20\% of the antineutrinos have oscillated away --- much more than found in any other experiment. The next curve to examine is the dot-dash (red) curve which is generated by $\chi^2_{mag}$, Eq.~\ref{eq6}, the magnitude (or rate) chisquare. This is the most commonly used chisquare function for analyzing the reactor anomaly experiments. First, we see that without the spectral information,  it has no sensitivity to a particular mass and is nearly a straight line. Secondly, it underestimates the significance of the experiment substantially; any analysis that uses the rate approach for an experiment that has spectral information will be significantly underestimating the impact of that experiment. Next we examine the results obtained from the spectral chisquare, Eq. ~\ref{eq7}, the dash (blue) curve. It too produces predictions of possible mass-squared differences, in fact, nearly identical values to those predicted by the conventional chisquare. The dot-dot-dash (indigo) curve is for the sum of the magnitude and spectral chisquares. It gives results that are reasonably close to the conventional chisquare. This supports our definition of the spectral chisquare. Finally the dot-dot-dash green curve is the result of the Mention spectral chisquare, Eq.\protect\ref{eq8}. These results are quite small. This is not surprising as the systematic errors are included twice.

The spectral chisquare, which is independent of the magnitude of the flux is of special interest. Note that because the Huber flux and the Daya Bay flux differ \cite{dbb} only in magnitude, the spectral chisquare, $ \chi^2_{spec}$, the dash (blue) curve, gives identical results for these two fluxes. The revision of an increase by 18\% of the flux appeared fourteen years after the original publication and is authored by a fraction of the original collaboration. It is also a much larger disappearance fraction than any other oscillation experiment. This makes us cautious of this change in the flux.  We see that the spectral chisquare produces results with the location of the valleys, best fit values, very similar to what was found from the full conventional chisquare with the minima reduced, but much deeper than that found by Mention \cite{Ment}.  If the flux increase is less than the full 18\% increase, the results will lie between the conventional chisquare solid (black) curve and the spectral chisquare dash (blue) curve. 

\section{Results - Flux dependence}

\begin{figure}
\includegraphics[width=4.2in]{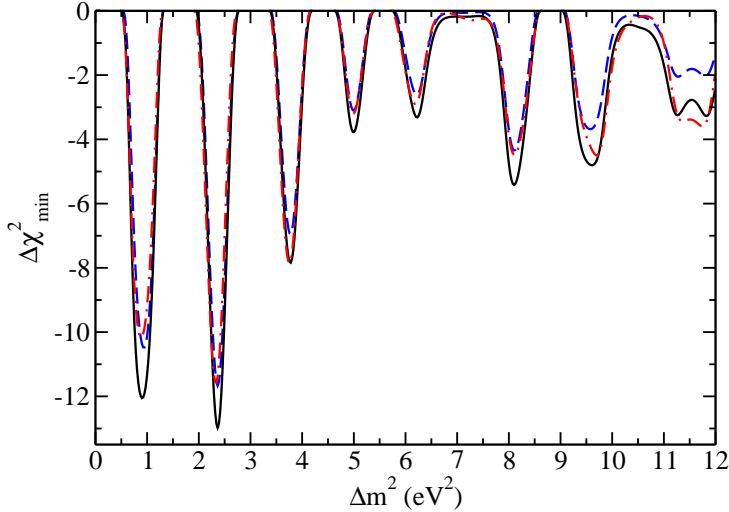}
\caption{$\Delta\chi^2_{min}$ versus $\Delta m^2$ for the conventional chisquare and three different fluxes. The solid (black) curve is for the Huber flux, the dash (blue) curve is for the Daya Bay flux, and the dot-dash (red) curve is for the ILL flux.}
\label{fig2}
\end{figure}

The question of the flux, both its magnitude and its energy dependence, has received much attention \cite{Hayed,Haya,Hub2,sonz,Hayee,dba,Hubb,dbb,Hayef} lately. The historical way of modeling the flux is to start with a measured beta decay spectrum and then theoretically predict a neutrino spectrum that is consistent with the measured beta spectrum. The most recent flux of this type is that given by Huber \cite{Hube}. The alternative is to measure the flux directly, the most recent such flux is given by the Daya Bay \cite{dba,dbb} collaboration. These two fluxes are not consistent with each other. The energy dependence of the flux for the Daya Bay measurement has a bump in the flux near 5 MeV that is absent in the Huber flux. The recent NEOS \cite{NEOS} experiment measures the flux for its particular mix of isotopes and finds corroberating evidence for this bump. The two approaches also do not agree on the magnitude of the flux. The Daya Bay  experiment sees a lower flux rate for its particular mix of isotopes than is predicted by Huber. It cannot tell you directly how much of the decrease comes from which isotope. Unfolding the decrease must be done theoretically. In Ref.~\cite{dbb}, the conclusion reached by the Daya Bay experimentalists is that the Daya Bay flux is a reduction by 7.8\% for the $^{235}$U flux with the other isotopes unchanged as compared to the Huber flux. We here present results, Fig.~\ref{fig2} and Table~\ref{tab2},  for the ILL experiment utilizing the Huber flux, the Daya Bay flux, and the ILL flux. We include the historical LL flux purely out of curiosity concerning what would have been the results had there been an analysis performed looking for a fourth neutrino, rather than focusing on the 90\% disallowed region, the general approach adopted at the time.
\begin{table}
\begin{center}
\begin{tabular}{|c|c|c|c|c|}
\hline
Flux&~~$\sin^2 2\theta_{min}$~~&~~$\Delta m^2_{min}$ (eV$^2$)~~&~~$\Delta\chi^2_{min}$~~&~~$\sigma$~~\\ 
\hline
\multirow{8}{*}{~ILL~}&$0.243$&$0.88$&$-10.2$&3.2\\ \cline{2-5}
                    &$0.251$&$2.34$&$-11.6$&3.4\\ \cline{2-5}
                    &$0.201$&$3.75$&$-5.38$&2.3\\ \cline{2-5}
                    &$0.223$&$3.73$&$-7.79$&1.2\\ \cline{2-5}
                    &$0.117$&$6.18$&$-1.09$&1.0\\ \cline{2-5}
                    &$0.181$&$8.12$&$-1.99$&1.4\\ \cline{2-5}
                    &$0.223$&$9.68$&$-1.64$&1.3\\ \cline{2-5}
                    &$0.181$&$11.6$&$-0.74$&0.9\\ \hline
\multirow{8}{*}{~Daya Bay~}&$0.239$&$0.95$&$-10.5$&3.2\\ \cline{2-5}
                    &$0.259$&$2.36$&$-11.7$&3.4\\ \cline{2-5}
                    &$0.213$&$3.78$&$-6.95$&2.6\\ \cline{2-5}
                    &$0.157$&$5.00$&$-3.10$&1.8\\ \cline{2-5}
                    &$0.165$&$6.23$&$-2.61$&1.6\\ \cline{2-5}
                    &$0.241$&$8.10$&$-4.34$&2.1\\ \cline{2-5}
                    &$0.243$&$9.56$&$-3.68$&1.9\\ \cline{2-5}
                    &$0.245$&$11.3$&$-2.05$&1.4\\ \hline
\end{tabular}
\caption{Results for the value of the minima of $\chi^2_{min}$ and their location, $\sin^22\theta_{min}$ and $\Delta m^2_{min}$ for the conventional chisquare, $\chi^2_{conv}$ and two fluxes, the ILL flux and the Daya Bay Flux. Results for the Huber flux is given in Table~\ref{tab1}.} 
\label{tab2}
\end{center}
\end{table}

The Huber flux for $^{235}$U is given in Appendix B of Ref.~\cite{Hube}. Rather than utilize the magnitude of the flux given there, we put an emphasis on staying as close to what the experimentalists did in their analysis as is possible. In the second ILL paper \cite{houm} and in the Mention paper \cite{Ment} we are given the ratio of the total number of experimentally measured neutrinos to the no-oscillation expected number, 0.802. In \cite{kwon} we find that the total number of electron antineutrinos measured is 4890. Thus the Mueller flux is to be normed to 6070 events. In \cite{Hube} we find the $^{235}$U Huber flux is 1.004  times the Mueller flux  or  is to be normed to  6100. From \cite{dbb} the Daya Bay flux is 7.8\% smaller than the Huber flux or is to be normed to 5620 counts. From \cite{Ment} the ILL flux is  2.6\%  smaller than the Mueller flux or is to be normed to 5910 and approximately has the energy dependence of the Mueller flux, which is given in Ref.~\cite{Mue}

$\Delta\chi^2_{min}$ versus $\Delta m^2$ is presented in Fig.~\ref{fig2}  for the Huber flux, the Daya Bay flux, and the ILL flux and for the conventional chisquare, $\Delta\chi^2_{conv}$. In addition, in Table~\ref{tab2} the depth of each $\Delta\chi^2_{min}$ and the location of the chisquare minima, $\sin^22\theta_{min}$ and $\Delta m^2_{min}$, are given for the Daya Bay and ILL flux. The results for the Huber flux was given in Table~\ref{tab1}.. We see that the change in the flux does not cause much of a change in the location of the $\chi^2_{min}$ and does not cause a major change in the depth of the $\chi^2_{min}$. This is because of the 20\% disappearance of the antineutrinos. This is sufficiently large that the 7.8\% reduction in the flux reduces the impact of the experiment, but not overwhelmingly. If we investigate an experiment where we have pure $^{235}$U fuel, and the Huber flux gave a 6\% or less disappearance, the reduced flux of the Daya Bay experiment would lead to a null result for the existence of a fourth neutrino. 

We see that all three fluxes give substantial evidence for the existence of a fourth neutrino. Indeed, the conventional chisquare implies the lowest two minima for the Daya Bay flux are quite deep, with $\Delta\chi^2_{min}$ given by -10.5 and -11.7  (3.2 and 3.4 $\sigma$). We see similarly that the ILL flux gives -10.2 and -11.6 (3.2 and 3.4 $\sigma$) for the depth of the two deepest minima. Had the ILL experiment been modeled with a conventional chisquare, the reactor anomaly would have been discovered much earlier.

\section{Origin of multiple minima}
\begin{figure}
\includegraphics[width=4.2in]{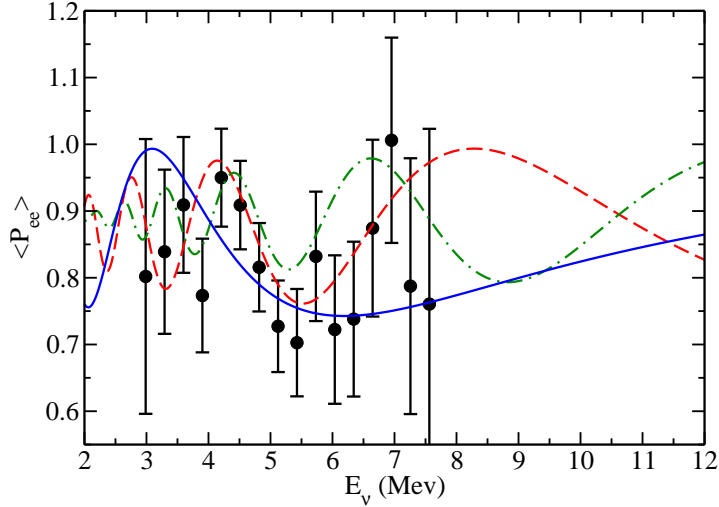}
\caption{The averaged oscillation probability $<{\mathcal P}_{ee}>$ versus the antineutrino energy. The data are from Ref.~\cite{houm}. The solid (blue) curve is the theoretical calculation for the first minimum in the chisquare; the dash (red) curve the second minimum; and the dot-dash (green) curve the third minimum.}
\label{fig3}
\end{figure}

Finding multiple minima brings up the question of whether the results are predicting more than one sterile antineutrino or are offering several possible values for the mass-square difference. The analysis was performed using an oscillation probability from a 3+1 model. Logically the results could not be for multiple sterile  antineutrinos. In Fig.~\ref{fig3}, the solid (blue) curve is for the first minimum of the chisquare function found at $\sin^2 2\theta = 0.26$ and $E_\nu = 6.0$ MeV. The dash (red) curve is for the second minimum found at $\sin^2 2\theta = 0.25$ and $E_\nu = 5.6$ MeV, and the dot-dash (green) curve is for the third minimum found at $\sin^2 2\theta = 0.22$ and $E_\nu = 5.3$ MeV.  All three curves have a minimum near 0.5 eV$^2$. What is happening is that the solid (blue) curve fits with its first minimum near 0.5 eV$^2$, the dash (red) curve fits with its second minimum near 0.5 eV$^2$, and the third curve fits with its third minimum near 0.5 eV$^2$. With less than perfect data, a fundamental and its harmonics can all produce reasonable fits. Thus the data is producing a series of possible mass-square differences. For data from a model calculation with small errors given in Ref.~\cite{conr},  it is shown how the data can distinguish between multiple possible single antineutrino solutions and a solution that actually represents the existence of multiple sterile antineutrinos.
One can be cautious of the suggested 18\% increase in the flux suggested in Ref.~\cite{houm}, but we believe that unless and until other data contradict this claim, the results of a full analysis of the reactor antineutrino data utilizing the standard chisquare should be the default.

\section{Conclusions}
Of the nineteen reactor anomaly experiments, the ILL  experiment has the shortest baseline, 8.78 m. In Ref.~\cite{houm} a correction to the measured power of the reactor during the experiment was reported and an increase by 18\% to the reactor flux was proposed. This means that approximately 20\% of the electron antineutrinos emitted from the reactor had oscillated away. This large fraction of antineutrinos disappearing would intuitively imply the existence of a sterile fourth antineutrino at the mass-squared scale $\Delta m^2 \ge$ 1 eV$^2$ and with a large probability for the existence of this sterile antineutrino. The analysis performed in Ref.~\cite{houm}, however, used an unusual and peculiar functional form for the chisquare  function, which we ignore. The analysis done in Ref.~\cite{Ment}, the work that originally proposed the existence of a reactor anomaly, used a spectral chisquare which we believe included the systematic errors twice. Other global analyses, Refs.~\cite{Kopp,Dent2,Coll,Gari,Dent}, either omitted the ILL experiment or used the magnitude chisquare, which we find underestimates the significance of an experiment that contains spectral information. We also demonstrate that the conventional chisquare, $\Delta\chi^2_{conv}$, can be quantitatively broken into a magnitude (or rate) part and a spectral part, with the spectral part, $\Delta\chi^2_{spec}$, given by the form that we propose in Eq.~\ref{eq7}

We find that using the standard, rigorously justified by mathematicians for normal statistics, chisquare function, $\Delta\chi^2_{conv}$, Eq.~\ref{eq1}, gives results that imply the existence of a fourth neutrino at a number of specific values for the possible mass-squared differences. The set of mass-squared differences preferred is given in Table I together with the statistical significance of each.  We also examine the results implied by the spectral chisquare, $\Delta\chi^2_{spec}$, given in Eq.~\ref{eq7}. The significance of an experiment is necessarily reduced by utilizing only the spectral form of the chisquare function, but there is the advantage of the results being independent of the magnitude of the flux.  We find for the Huber flux that $\Delta\chi^2_{min}$ for the two lowest mass-square differences are  -12.1 and -13.0 (3.5 and 3.6 $\sigma$ ) with mass-squared differences of 0.90 and 2.36 eV$^2$. For the spectral chisquare,  $\chi^2_{spect}$ the mass-squared difference values of the minima remain nearly the same as those found for the conventional chisquare, 0.95 and 2.36 eV$^2$,and have a depth of  -8.22 and -9.45 (2.9 and 3.1 $\sigma$)..  We note that the spectral chisquare puts a lower limit on the implications of an experiment that can result from not knowing the magnitude of the flux. The value for the magnitude chisquare, $\Delta\chi^2_{mag}$, for the Huber flux is found to be -4.0 (2.0$\sigma$) and independent of the value of $\Delta m^2$ for $\Delta m^2 > 0.1$ eV$^2$.

We find that the use of the magnitude chisquare (rate analysis) underestimates the significance of an experiment that has spectral information. Studies of the reactor anomaly experiments, with the exception of the Daya Bay experiment, utilize a rate analysis or ignore the ILL experiment. This has motivated us to redo all nineteen experiments in which we will include this new analysis of the ILL experiment and spectral information when available. We also find that when spectral information is included, each experiment  predicts individual values for $\Delta m^2$ that are preferred. This alters how one can view the process of combining individual experiments. The question of coherence between the individual values preferred by one experiment and those values found by all the other experiments becomes very important. The discussion of coherence between the $\Delta m^2$ values found here for the ILL experiment and the values found by other experiments will be presented when the new results for the reactor anomaly are complete. In addition  there are five newer reactor anomaly experiments that have been published or have preprints that have appeared in the archive. These also need to be combined and included in with the older experiments. These experiments are Nucifer \cite{Nuc}, NEOS \cite{NEOS}, Nuetrino-4 \cite{N-4}, DANSS \cite{DANS}, and PROSPECT \cite{PROS}. As these experiments should be more reliable than the older experiments, the question of coherence becomes a more important consideration.

The question of the magnitude of the flux remains. With 20\% of the antineutrinos disappearing, the ILL experiment finds that the 7.8\% reduction for the Daya Bay flux does reduce the impact of the ILL experiment, but leaves the results with the deepest two values of $\Delta\chi^2_{min}$ at the significant values of -10.5(3.2) and -11.7(3.4). The PROSPECT experiment will measure the $^{235}$U flux to a much improved accuracy, both its measured energy dependence and its magnitude.  For research reactors that use pure $^{235}$U, the flux question will be resolved. For other experiments, this measurement will reduce the uncertainty.  If the Daya Bay flux is confirmed, the question will be resolved. Otherwise, given the level of discussion, Refs. \cite{Hayed,Haya,Hub2,sonz,Hayee,dba,Hubb,dbb,Hayef}, we will follow the conclusion of Ref.~\cite{Hayef}, ``The present analysis suggests that there is currently insufficient evidence to draw any conclusions on this issue." Further measurements would be necessary.

\end{document}